\documentclass[epj,draft]{svjour}
\usepackage[]{psfig}
\begin{document}
\title{Asymmetric Mach-Zehnder fiber interferometer test of the anisotropy of the speed of light}
\author{Victor de Haan}
\institute{BonPhysics B.V., Laan van Heemstede 38, 3297 AJ Puttershoek, The Netherlands, \email{victor@bonphysics.nl}}
\date{Version: August 2009}
\abstract{
Two optical fiber Mach-Zehnder interferometers were constructed in an environment with a temperature stabilization of better than 1~mK per day. One interferometer with a length of 2~m optical fiber in each arm with the main direction of the arms parallel to each other. A path (length 175 mm) filled with atmospheric air is inserted in one arm. Another interferometer with a length of 2~m optical fiber in each parallel arm acts as a control. In each arm 1~m of fiber was wound around a ring made of piezo material enabling the control of the length of the arms by means of a voltage. 
The influence of rotation of the interferometers at the Earth surface on the observed phase differences was determined. For one interferometer (with the air path) it was found that the phase difference depends on the azimuth of the interferometer. For the other one no relevant dependence on the azimuth has been measured.   
}

\maketitle
\section{Introduction}
At the end of the 19th century is was generally believed that first order experiments, where the effect depends in first order on the ratio $v/c$, where $v$ is the velocity of the observer with respect to a preferred rest frame~\footnote{This could be the frame defined by Newtons conception of absolute space, a distinguished frame of reference relative to which bodies could be said to be truly moving or truly at rest. A candidate for this is the Cosmic Microwave Background Radiation frame~\cite{Smooth1977}. Otherwise it could be a local frame where light-speed anisotropy is due to the curvature of spacetime~\cite{DiSalle2008}.} and $c$ is the speed of light in this frame, could not be used to detect this velocity. An explanation of this was the famous Fresnel ether drag, which would compensate any preferred rest frame effect, at least to first order. Recently, it has been argued by Cahill~\cite{Cahill2003} and Consoli~\cite{Consoli2004} that the Miller effect~\cite{Miller1933}, together with all other Michelson Morley interferometer experiment results~\cite{Piccard1926,Piccard1928,Illingworth1927,Joos1930}, could be caused by an absent of ether drag. This drag would depend on the difference of the refractive index of 1, which for atmospheric air is approximately $3 \times 10^{-4}$, for atmospheric helium $4 \times 10^{-5}$ and for vacuum $0$. This would also explain why modern-day vacuum experiments all give much lower limits for the anisotropy. Trimmer~\cite{Trimmer1973} did just such a measurement where he took a Sagnac type of interferometer and inserted a piece of crown glass (refractive index 1.5, length 120~mm) in one of the beams. He deduced from his experiment an anisotropy of less than $10^{-10}$ for the first order and less than $2\times 10^{-11}$ for the third order. However, his experiment was performed in vacuum and was insensitive to second order effects.
\\
The present paper reports on an experiment where in one arm of a Mach-Zehnder interferometer a path of air was inserted introducing a possible difference in ether drag.
\\
\section{Outline of the method}
\begin{figure}[tbp]
\begin{picture}(250,250)
\put(0,0){\psfig{figure=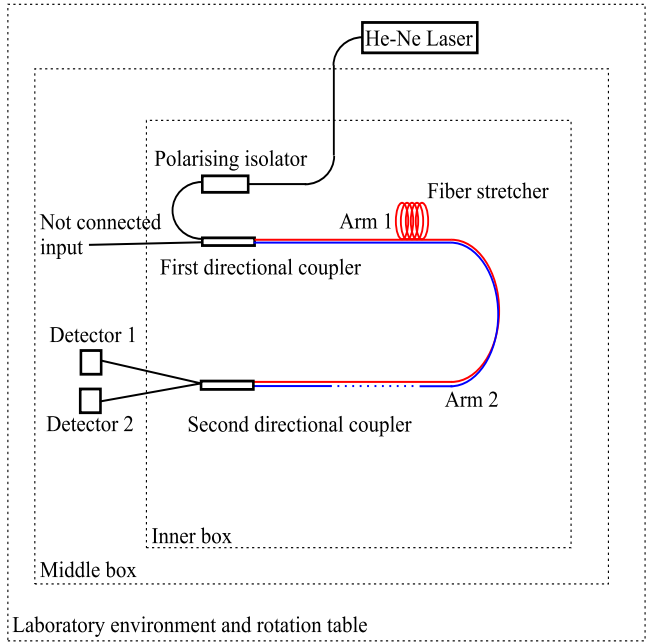,height=85mm,width=85mm}}
\end{picture}
\caption{\label{fig1} Configuration of the asymmetric Mach-Zehnder interferometer.}
\end{figure}

The method is similar to the one described in~\cite{DeHaan2009}. With an asymmetric fiber Mach-Zehnder interferometer (see figure~\ref{fig1}) the light injected into a fiber by means of a laser is split into two equal and in phase parts by a 2x2 directional coupler. These two light beams travel through two (parallel) fiber arms of the interferometer and are rejoined by a second 2x2 directional coupler. Depending on their mutual phase, the light beams interfere constructively for one output fiber and destructively for the other one or vice versa. The two outputs of the second 2x2 directional coupler are fed into two detectors where their intensities $I_1$ and $I_2$ are measured. 
The sum of the intensities is proportional to the laser output power. The difference of the two intensities relative to their sum is called the visibility, $V$. For an ideal interferometer the visibility would change between -1 and +1, depending on the phase difference, $\phi$ of the light beams in the second directional coupler according to
\begin{equation} \label{eq1}
 V = \frac{I_1-I_2}{I_1+I_2}=\cos{\phi} .
\end{equation}
The phase difference is determined by the optical path of the light, while traveling along the fibers from the first directional coupler to the second one. This optical path depends on the length of the arms and the wavelength of the light moving through the fibers of the arms. The wavelength, $\lambda$  of the light depends on the speed of light in the fibers and hence on the refractive index, $n$ of the fibers and (possibly) on the velocity and motion direction, $\vec{v}$ of the Earth relative to a preferred rest frame. Hence,
\begin{equation} \label{eq2}
\phi =\oint_{1} \frac{2\pi}{\lambda(n,\vec{v})}dl -  \oint_{2} \frac{2\pi}{\lambda(n,\vec{v})}dl ,
\end{equation}
where $\oint_{i}$ denotes the line integral along arm $i$ from the first directional coupler to the second one. The determination of the change in phase difference due to the motion of the Earth is the objective of this experiment. Further details on the measurement method are described in~\cite{DeHaan2009}.
\section{Experimental set-up}
\begin{figure}[bt]
\begin{picture}(250,140)
\put(0,0){\psfig{figure=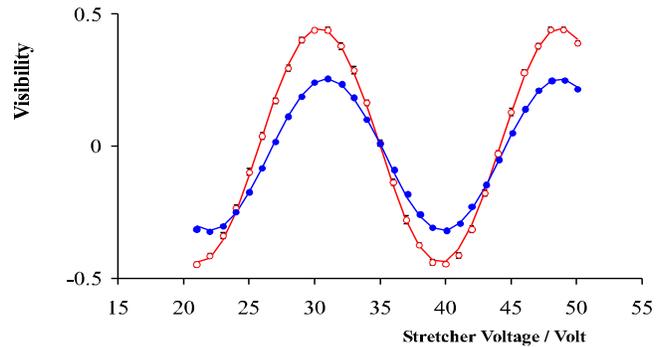,height=45mm,width=85mm}}
\end{picture}
\caption{\label{fig3} Visibility of both interferometers. (red) circles: {\it control} and (blue) dots {\it effect}) as function of the voltage applied to the fiber stretchers.}
\end{figure}

Two interferometers were coupled by means of an additional directional coupler to the same laser. One with its arms parallel to each other with a path of air inserted as shown in figure~\ref{fig1}. Another with its arms parallel without air as a control. Any anisotropy in the speed of light should turn up in the first one and should cancel in the second one. The length of the fibers in the arms of the effect interferometer was 2~m. The air path was 175 mm constructed by two fiberports (Thorlabs Inc. PAF-X-5) and a fiber table (Thorlabs Inc. FT-38X165). The length of the fibers in the arms of the control interferometer was 2~m. 
In one of the arms of the interferometers 1 meter long fibers were inserted wrapped around a thin-walled cylinder made of piezo material acting as a fiber stretcher. The typical response of the fiber stretchers as function of applied voltage is shown in figure~\ref{fig3}. The lines are cosine fits to the measured points with a frequency of 0.329(2)~rad/V and an amplitude of 0.393(2) for the stretcher in the {\it control} interferometer and 0.329(2)~rad/V respectively 0.363(4) for the other one. The frequency corresponds to a change in the circumference of the cylinders of about 18~nm/V. The standard deviation in the measured visibility is almost as large as the symbols. Note that the visibility amplitude of both interferometers is quite different and much smaller than 1. Part of it is due to the reduced coherence of the laser light after joining in the coupler as the difference in path is 1.00~m for the control and 1.18 m for the effect interferometer. It could also be due to transmission or polarization effects as discussed in~\cite{DeHaan2009}.
The temperature control was the same as described in~\cite{DeHaan2009}.
\begin{figure}[tb]
\begin{picture}(250,390)
\put(8,270){\psfig{figure=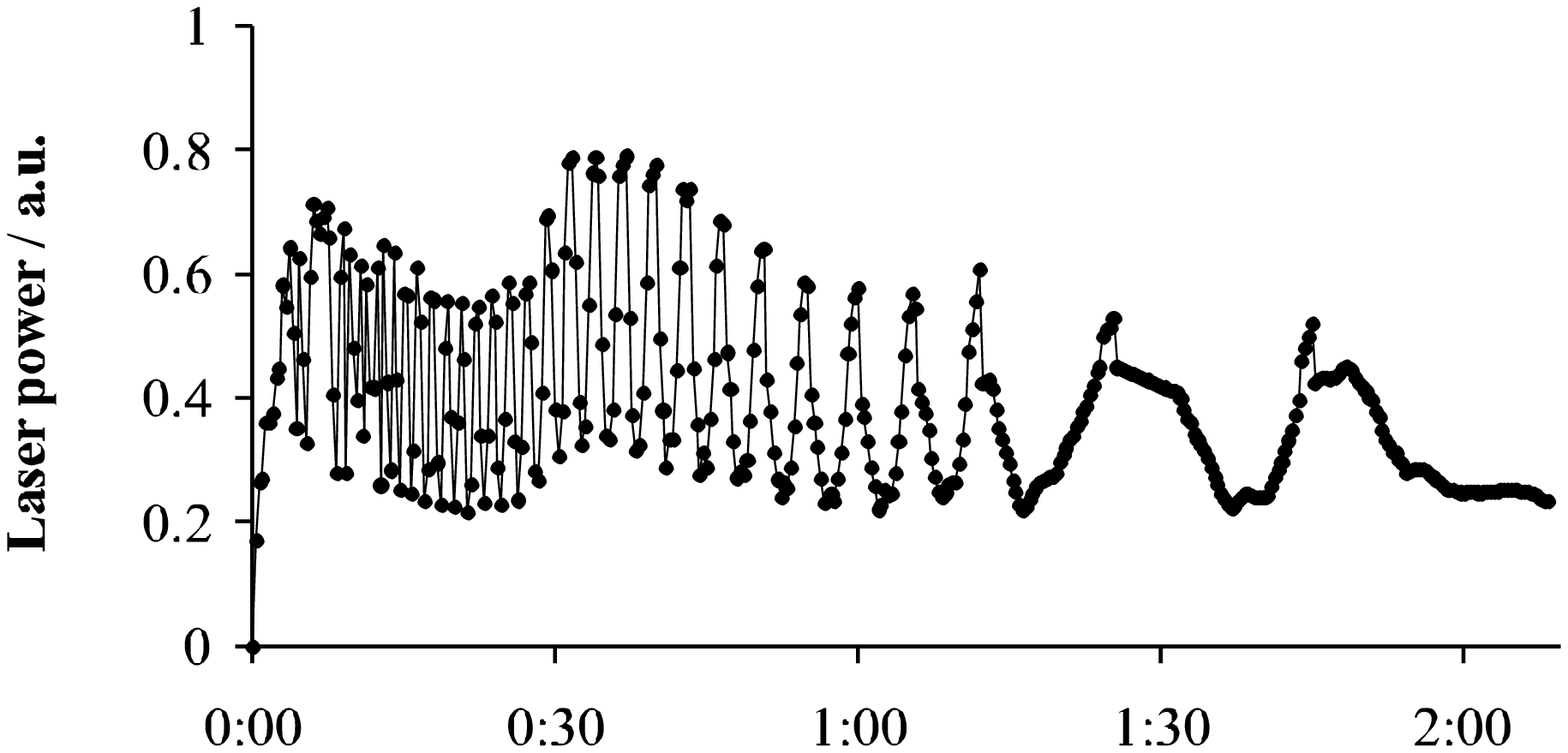,height=45mm,width=80mm}}
\put(6,135){\psfig{figure=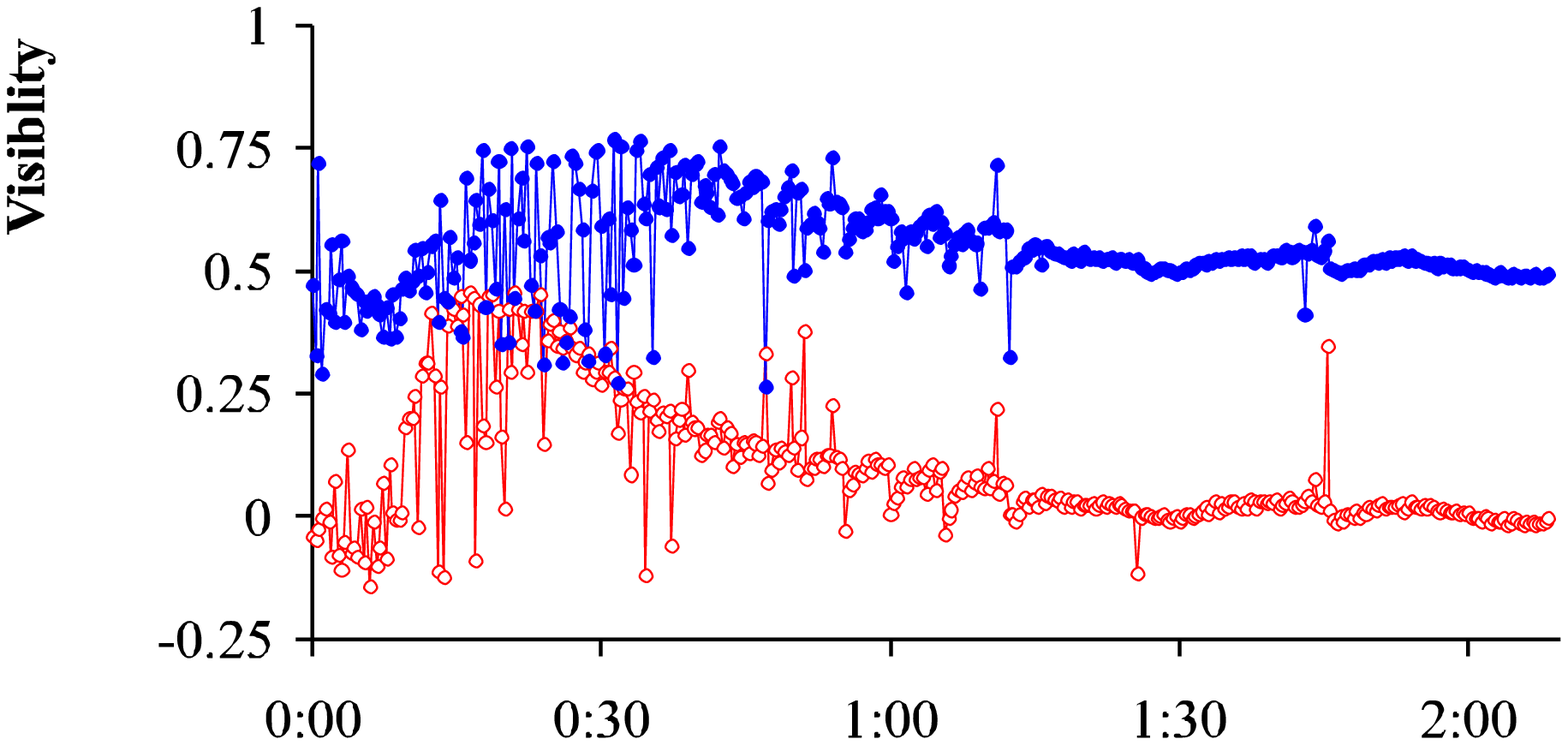,height=45mm,width=80mm}}
\put(2,2){\psfig{figure=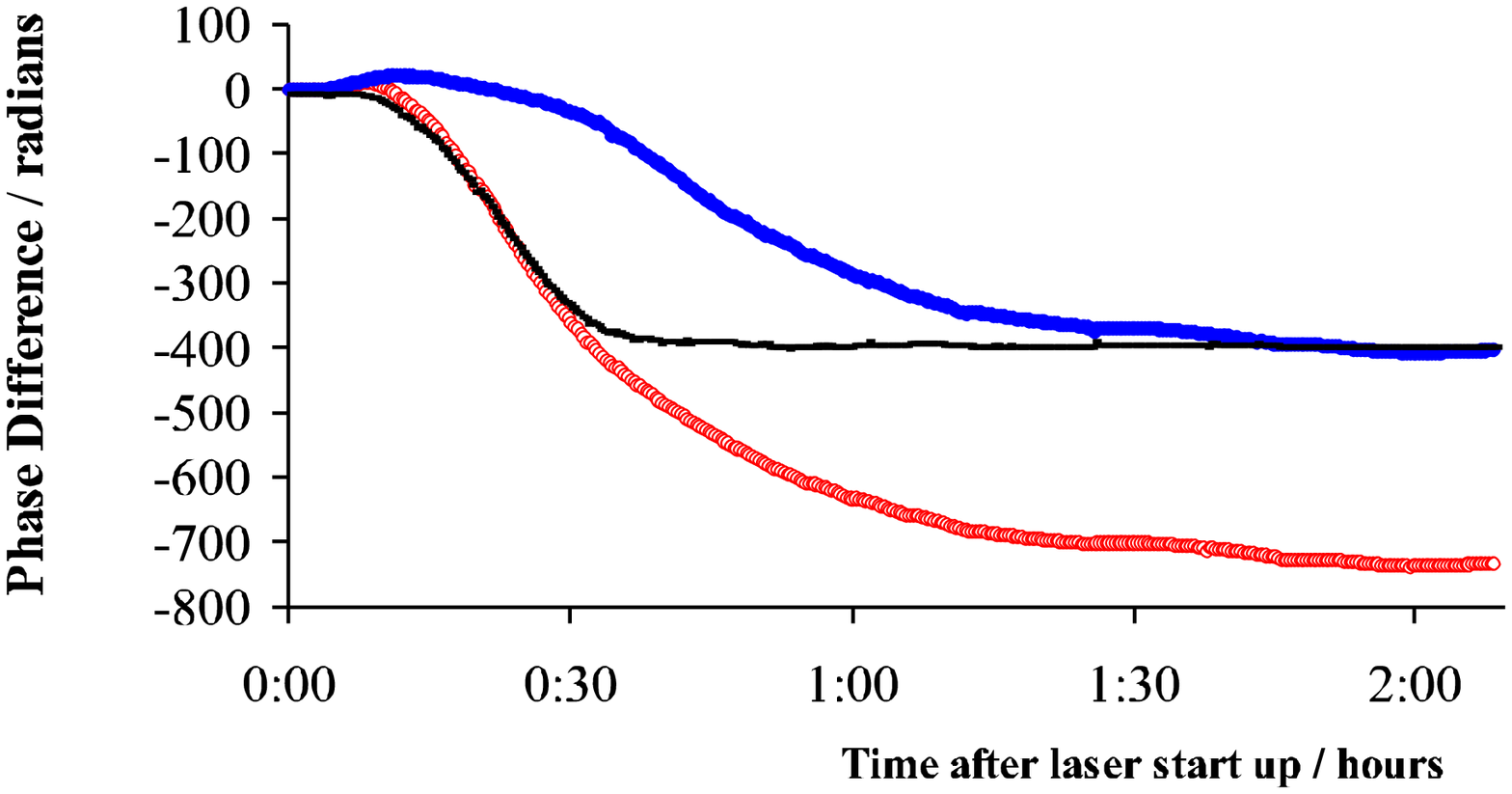,height=45mm,width=80mm}}
\end{picture}
\caption{\label{fig3B} Top: laser power as function of warm up time showing mode changes as changing intensities. Middle: Visibility of both interferometers (blue {\it control} and red {\it effect}, shifted with +0.5) during the same period. Bottom: Phase difference of both interferometers during the same period.}
\end{figure}
To find the effect of the laser frequency on the phase difference of the interferometers, the laser was switched off and switched on again after a cool down period of one hour. It takes the laser some time to become stabilized again. During this warm up period the laser experiences regular mode changes, depending on the size of laser cavity. These mode changes of 0.565~GHz change the phase difference of the interferometers and are shown in figure~\ref{fig3B}. The laser was turned on and after 2 minutes the phase control was turned on. The visibilities then rapidly decrease to around 0 and the phase difference is controlled by the stretchers. The visibility then becomes an error signal, indicating the accuracy of the controlled phase. Due to a time constant of the control of 1 s, rapid changes in phase difference create an error signal, which is used to change the voltage applied to the shifters. From this graph it is clear that both interferometers are extremely sensitive to a change in laser frequency. Even after a long time still some deviations occur. The extreme sensitivity is due to the length difference of the arms of the interferometer. The difference is at least 1 meter of fiber, increasing the effects described in~\cite{DeHaan2009} by at least a factor of 100. However, the ration between the phase difference of the control interferometer and the effect interferometer is constant (0.85). The black line shows the signal created by taking 0.85 times the phase difference of the control interferometer minus the phase difference of the effect interferometer. As long as the laser is stabilized the mode changes are less than 10~MHz, and the above signal reduce the effect to negligible proportions.
\\
The influence of the temperature on the phase difference has been determined by varying the temperature of the most inner container as a cosine with an amplitude of 0.02 K and a period of 6 hours. The results are shown in figure~\ref{fig4}. 
\begin{figure}[tb]
\begin{picture}(250,260)
\put(0,130){\psfig{figure=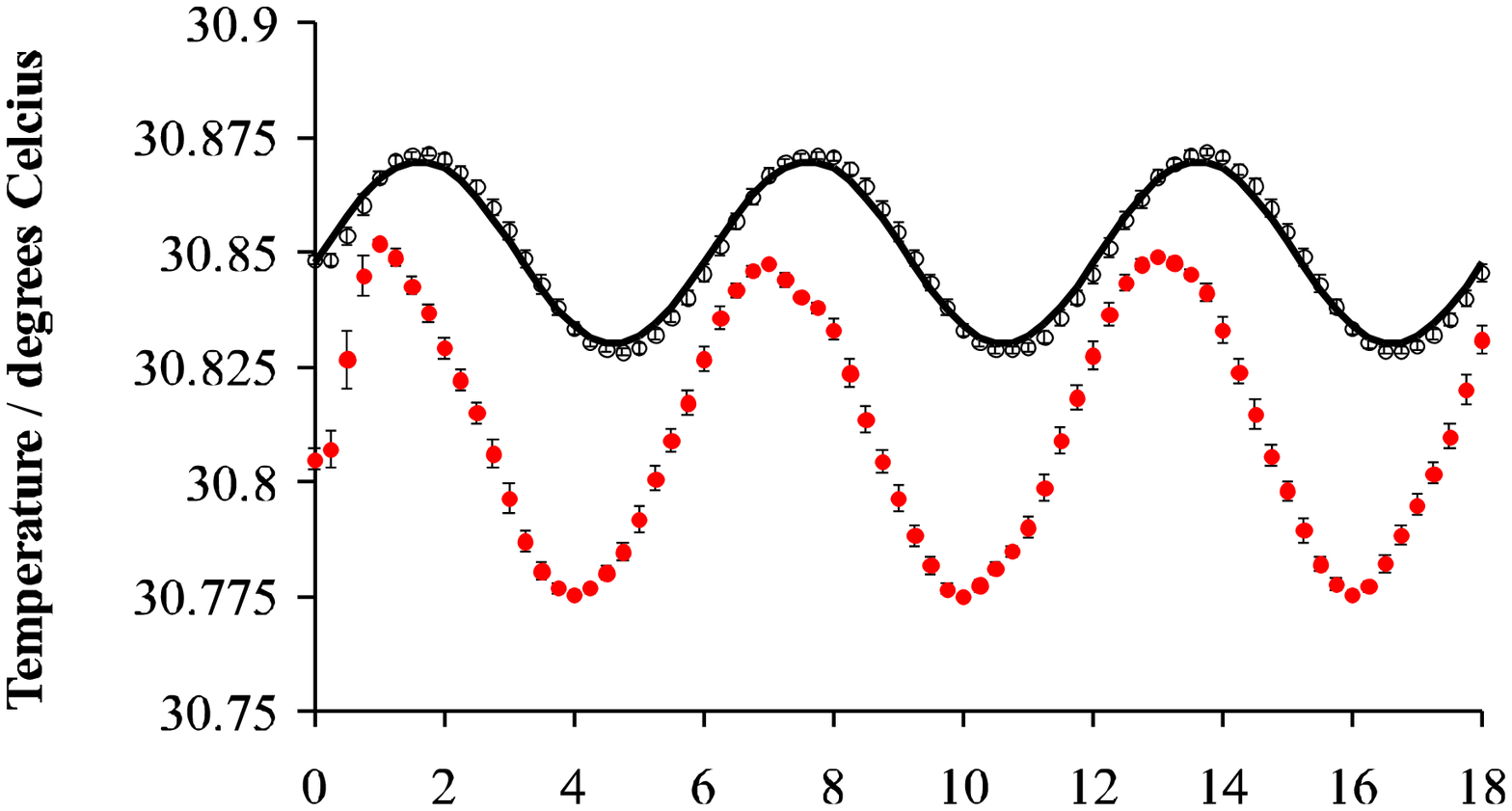,height=45mm,width=80mm}}
\put(4,2){\psfig{figure=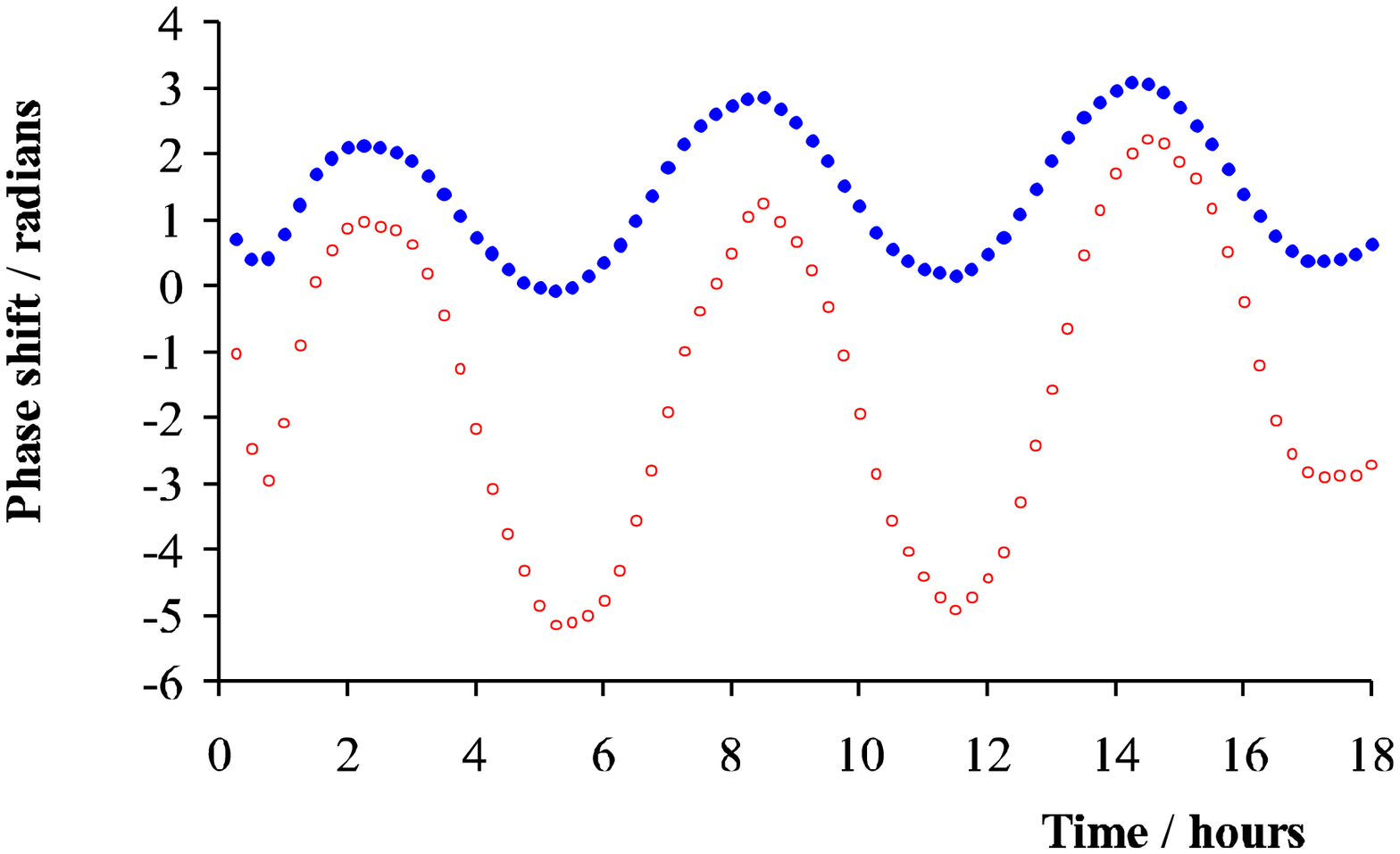,height=45mm,width=80mm}}
\end{picture}
\caption{\label{fig4} Top: set-point (black line), control temperature (black circles) and air temperature (red dots) as function of time. Bottom: phase difference of interferometers: (red) circles {\it effect} and (blue) dots {\it control} during the same period as the upper graph. The phase oscillations due to the temperature oscillations are reproduced imposed on a linear drift of unknown origin.}
\end{figure}
The black line in the top graph shows the set-point temperature. The black circles and error bars represent the control temperature. The deviations are due to the limitations of the temperature control. The red dots represent the temperature of the air inside the most inner box during the same time. At the start of the horizontal scale one can see the initial response of the temperature on the change in the set-point. Note that the amplitude is larger and it reaches a maximum before the set-point temperature. This gives an indication that during these temperature effect measurements the temperature distribution in the inner box is not stable. The curves in the bottom graph show the corresponding oscillations of the phase difference of the {\it effect} interferometer (in red circles) and for the other one (in blue dots). The oscillations correspond to a sensitivity to temperature variations of 150~rad/K for the {\it effect} interferometer and for the other one to a sensitivity of 70~rad/K. These values are of the same order as can be estimated from the properties of the used fibers~\cite{DeHaan2009}. The difference between the two sensitivities is for only a small part due to the air path in the {\it effect} interferometer as will be shown later. An extra linear decrease or increase in the phases as function of elapsed time is also observed. Here, for the {\it effect} interferometer, this change is somewhat larger than for the {\it control} interferometer. This is typical for all experiments done with these interferometers. 
\\
The influence of the atmospheric pressure has been estimated to be of the order of 1~rad/mbar for 1~m of fiber~\cite{DeHaan2009}. The influence on the refraction index of air can be found by using Edlen~\cite{Edlen1967}:
\begin{equation} \label{eqnair}
(n(p,T)-1)\times 10^{8}=\frac{1.04126p}{1+0.003671(T-273.15)}\times
\end{equation}
\[
\left\{8342.13+\frac{2.406030\lambda^2}{130\lambda^2-1000000}+\frac{0.015997\lambda^2}{38.9\lambda^2-1000000}\right\},
\]
where $p$ is the pressure in bar, $T$ the temperature in Kelvin and $\lambda$ the wavelength in nanometer. Hence, the phase difference change in an interferometer arm of $L=175$~mm length due to a change in temperature or pressure becomes according to eq.~\ref{eq2} for $\lambda=633$~nm:
\[
\Delta \phi \approx -1.5  \mbox{ rad/K} ,
\]
and
\[
\Delta \phi \approx 0.45 \mbox{ rad/mbar}.
\]
The atmospheric pressure was measured during the experiment with an accuracy of 0.05~mbar yielding a resulting error of 0.02~rad. The temperature was controlled to be constant within a standard deviation of 2~mK, yielding an absolute accuracy of 0.3~rad. It should be stressed that this is the standard deviation of the air temperature in the inner box during the complete measurement time of more then a month. Standard deviation for an hour measuring time were well below the measurement accuracy (0.5~mK).

\section{Results and discussion}
\begin{figure}[tb]
\begin{picture}(250,230)
\put(3,0){\psfig{figure=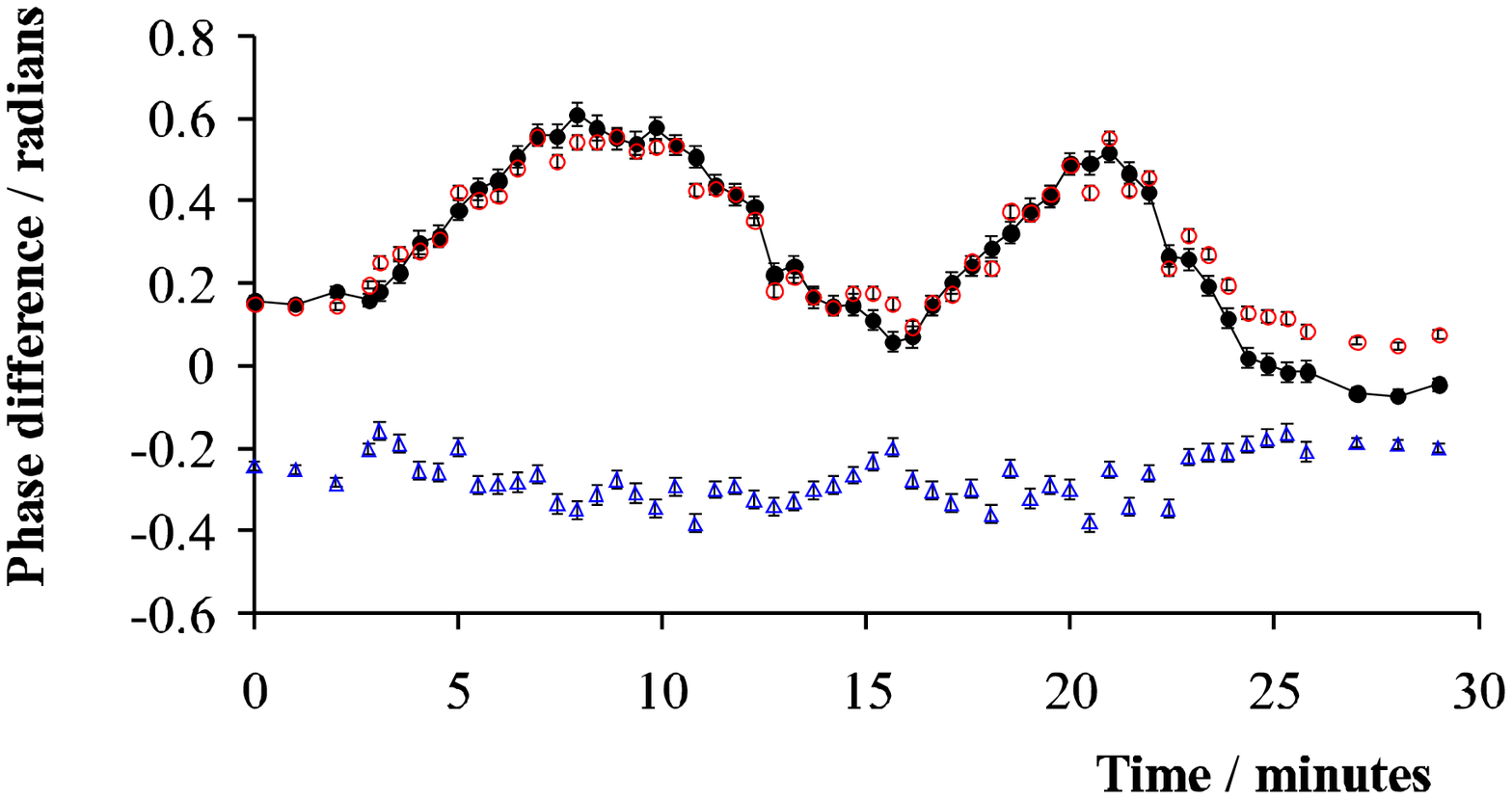,height=40mm,width=80mm}}
\put(3,115){\psfig{figure=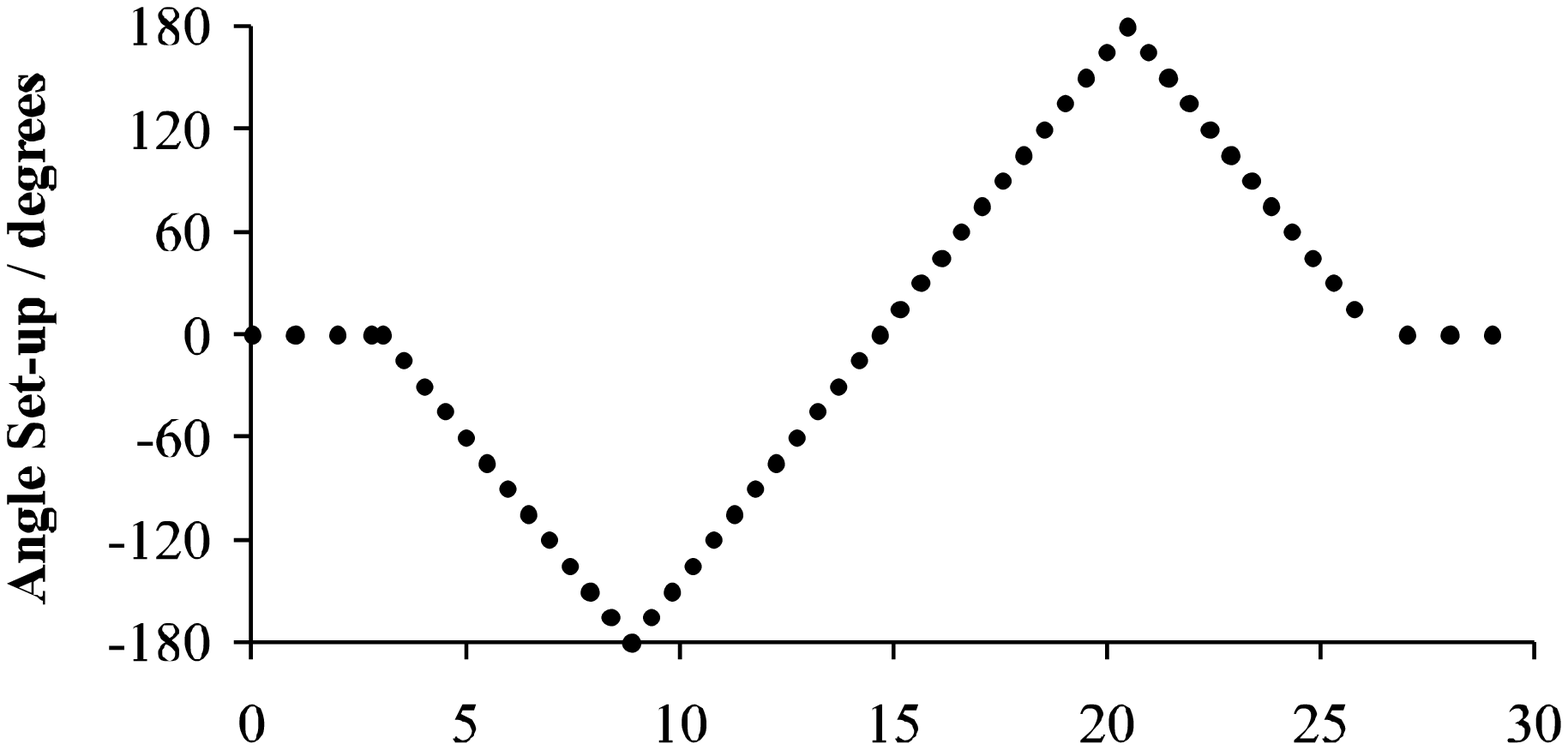,height=40mm,width=80mm}}
\end{picture}
\caption{\label{fig5} Top: angle of set-up as function of time. Bottom: open (red) circles: phase difference of {\it effect}; open (green) triangles {\it control} interferometer during the same period as the upper graph and closed (black) dots corrected phase difference again during the same period.}
\end{figure}

\begin{figure}[bt]
\begin{picture}(250,240)
\put(0,120){\psfig{figure=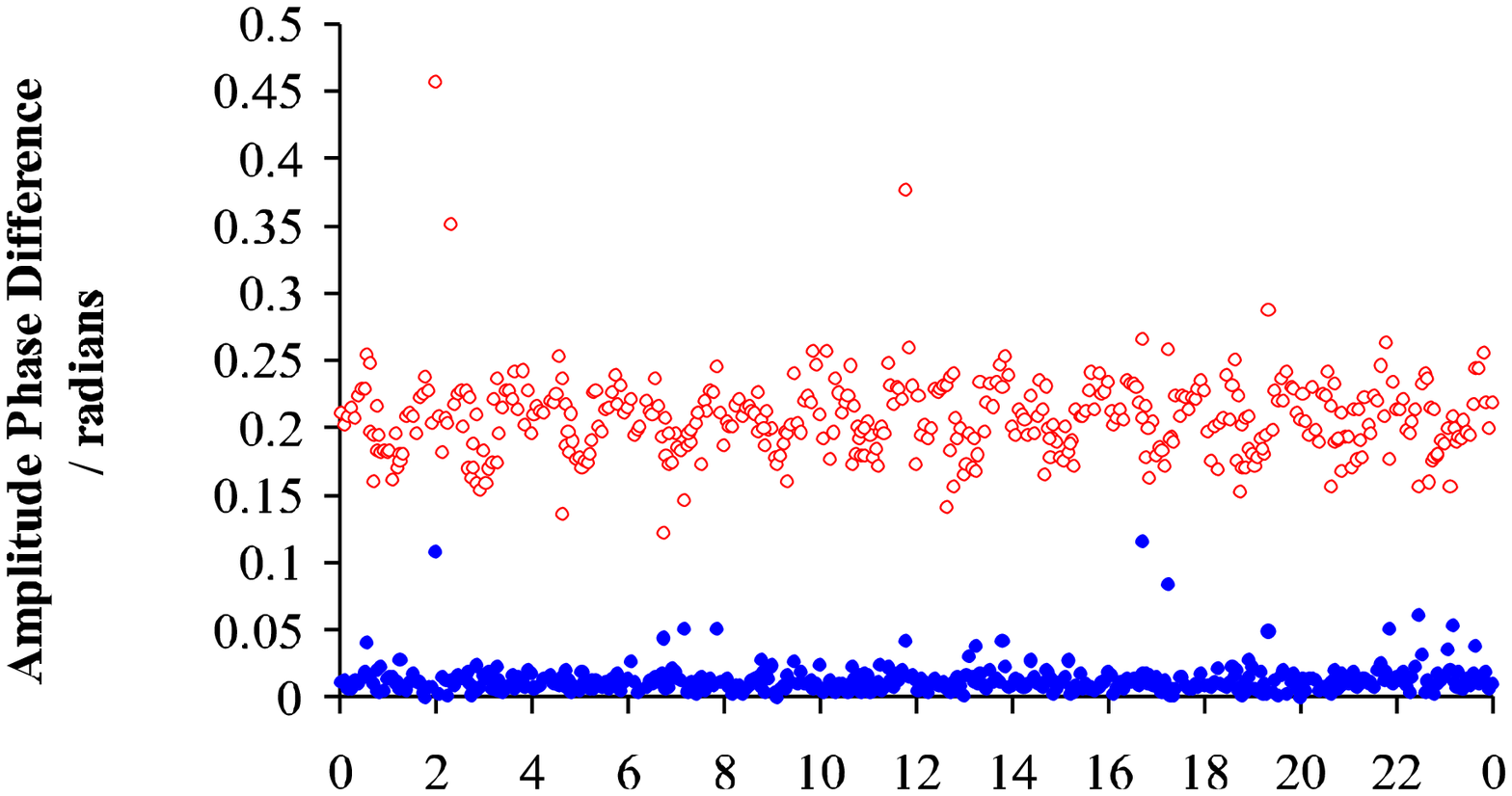,height=40mm,width=80mm}}
\put(4,0){\psfig{figure=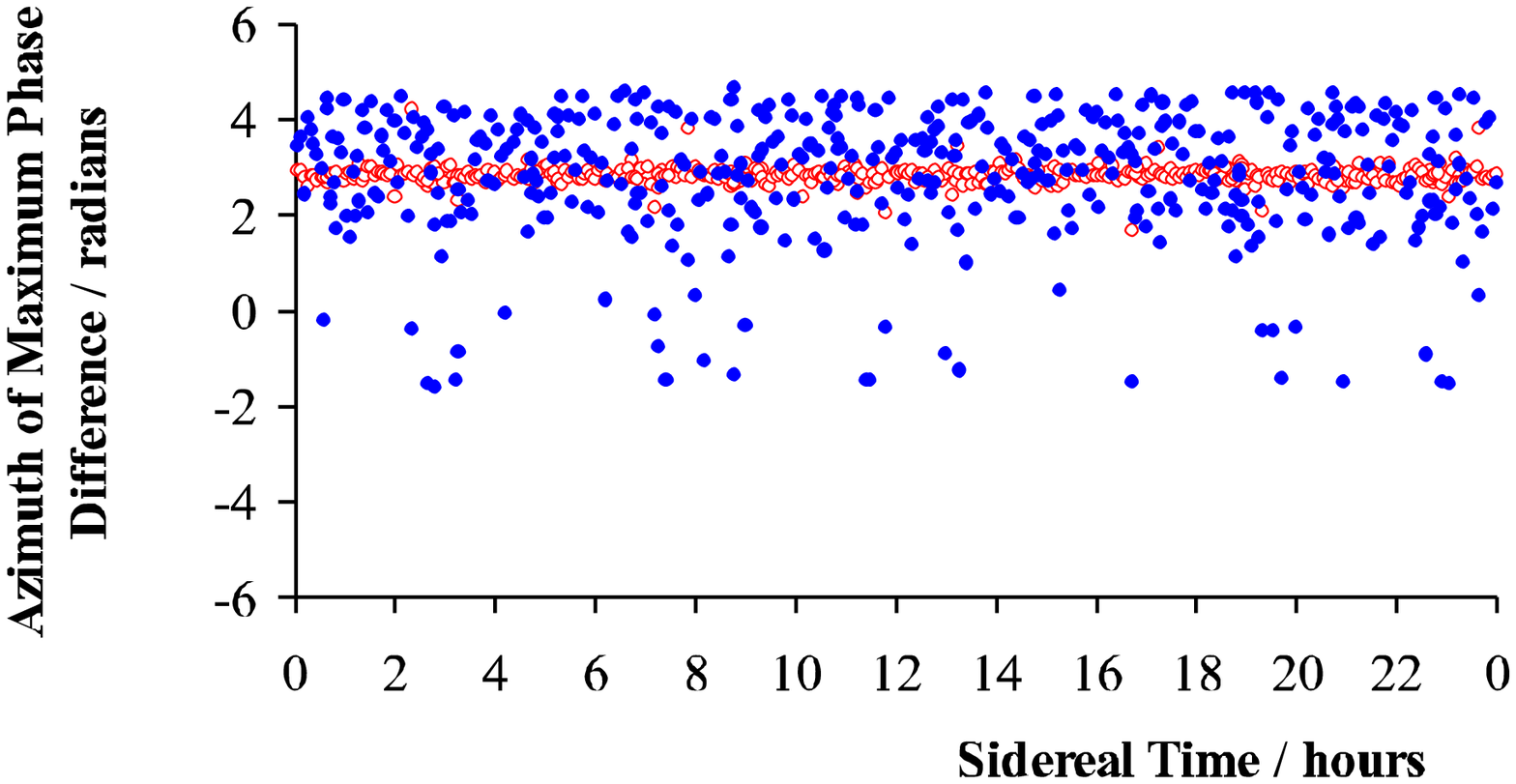,height=40mm,width=80mm}}
\end{picture}
\caption{\label{fig8ab} Top: first (red) and second (blue) order amplitude of the signal of the {\it effect} interferometer as function of sidereal time. Bottom: first (red) and second (blue) order azimuth of the maximum of the same signal. For clarity reason the error bars have been omitted. They have approximately the same length as the spread in points.}
\end{figure}

Finally, the set-up is rotated along a vertical axis to find out if there is an influence due to the velocity of the Earth in a preferred rest frame. Every 2 hours the set-up rotated from 0 to -180 degrees, from -180 to +180 and from +180 to 0 with steps of 15 degrees. For 0 degrees the path of air arms points to the local North. For 90 degrees it points to the local East. The longitude and latitude of the location of the interferometers on Earth are 4.7 degrees and 51.4 degrees. At every step the phase of both interferometers is measured during 15~seconds to average out statistical variations. A complete rotation takes about 20 minutes. A typical response of the interferometers as function of angle is shown in figure~\ref{fig5}.
The top graph of this figure shows the angle of the set-up as function of time. At the bottom (blue) triangles show the phase difference of the {\it control} interferometer during the same period as the upper graph. The  phase difference of the {\it effect} interferometer is shown as (red) circles. The (black) dots are the {\it effect} signal corrected by subtracting 0.85 times the {\it control} signal (and shifting it with a constant value to be able to compare it with the uncorrected {\it effect} signal). Note that the so corrected signal has less variation than the uncorrected signal. The amplitude of the signal is approximately 0.20~radians, larger than fluctuations due to temperature fluctuations. The change in the signal is clear and reproducible. In the graphs the phase oscillations due to the rotation of the set-up are clearly visible, again imposed on a drift of unknown origin. At first sight, it seems that this interferometer gives an indication that there might be some effect on the phase difference due to the motion of the Earth. However, if such an effect exists it should depend on the time of day and year. Upon rotation of the set up, the amplitude and azimuth of the maximum should vary between certain minima and maxima depending on the orientation of the Earth velocity with respect to the preferred frame as discussed by M\'{u}nera~\cite{Munera1998}. The Fourier transform of the phase difference (corrected for the linear assumed drift) as function of rotation angle gives this amplitude and azimuth of the maximum phase difference for all orders. The zeroth order is just the average phase difference during a rotation. The first order represents the amplitude and azimuth of that part of the signal that varies with the cosine of the angle of the set-up, corresponding to first order effects in $v/c$. The second order represents the amplitude and azimuth of that part of the signal that varies with the cosine of twice the angle of the set-up, corresponding to second order effects, and so on. The error in the values can be estimated from the difference between the Fourier transform of the data points measured for increasing set-up angles and the one measured for decreasing set-up angles (to find the systematic error due to the unknown drift) combined with the Fourier transform of the variances (to find an estimate of the statistical error). The amplitude and azimuth should vary with the sidereal time and epoch. 
\\
The amplitude and azimuth of the maximum measured as function of sidereal time in the period from June 16, 2009 and July 25, 2009 are shown in figure~\ref{fig8ab}. There is no apparent fluctuation of the signal depending on sidereal time in any of the graphs. Hence, the conclusion is justified that the built asymmetric Mach-Zehnder fiber interferometer is not able to measure a sidereal effect on the anisotropy of the velocity of light on the Earth surface.
\\
When the first order signal of the interferometer with the air path is interpreted in a similar way as was done by Trimmer~\cite{Trimmer1973}, the result would be that the first order amplitude of the signal is less than 0.20(2)~radians corresponding to 0.030(3) parts of a fringe. Using the simple formula for the amplitude of the signal $\Delta \phi=2\pi L/\lambda v/c$, derivable from Trimmer~\cite{Trimmer1973} the result is a maximum possible velocity of 34(4)~m/s. However, the validity of this simple formula is questionable. During the end of the 19th century and the start of the 20th century it was believed that {\it all} first order effects are completely compensated by the Fresnel drag. This was stressed during the conference on the Michelson Morley experiments held at the Mount Wilson Observatory in February 1927~\cite{Conference1927}. The reason for this must be sought for in the Doppler effect~\cite{Christov2006}, which renders the calculation of time differences useless when the wavelength of the light is not taken into account. If this is the case, the first order effects measured here are due to other influences. For instance, they can be due to phase shifts introduced by changing stresses during the rotation of the set-up. The set-up was carefully leveled, reducing the first order effect in the {\it control} interferometer as much as possible. Although this is an indication that the set-up is perfectly horizontal, it does not guarantee that stresses in the {\it effect} interferometer are also reduced to negligible proportions. 
\\
Consoli~\cite{Consoli2004} argues that Fresnel drag, together with Lorentz contraction, compensates first and second order effects for interferometers with arms in vacuum or with arms containing a transparent solid and/or liquid, but that for gaseous substances the Fresnel drag may be absent. The amplitude of the phase difference upon rotation due to the path of air is $\Delta \phi=2\pi L/\lambda(1-1/n_a^2) v/c$, where $n_a$ is the index of refraction of air (equation~(\ref{eqnair})). The maximum first order effect extracted from this is 64(6)~km/s, about twice the velocity of Earth in its orbit around the Sun. However, the azimuth of the maximum of the first order effect is in the North-South direction and does not depend on sidereal time, indicating that if the explanation of Consoli would be valid for the magnitude, it does not yield the direction in the experiment. This could be due to a large component of the velocity perpendicular to the orbital plane of the Earth around the Sun. Consoli's hypothesis could be tested further by exchanging the air by helium, reducing the effect by an order of magnitude. 
\\
The average amplitude of the second order effect in the measurements reported here is 0.012(2)~rad, resulting in a maximum velocity of 25(4)~km/s when using $\Delta \phi=2\pi L/\lambda (v/c)^2$. This is somewhat higher than the results obtained from all previous Michelson-Morley type interferometer experiments, but its comparable to the first order effect calculated according to Consoli. The upper limit of the third-order signal measured by Trimmer~\cite{Trimmer1973} is  $2\times 10^{-11}= k (v/c)^3$, where $k$ is of the order of $1$. Taking $k=1$ this indicates a maximum velocity of 80~km/s, well above the limit established here. 
\section{Conclusions}
The two optical fiber interferometers built in a temperature controlled environment enable the determination of frequency, temperature and pressure effects on the phase difference of the interferometer arms.
\\
Upon rotation of the interferometers around a vertical axis in one of the interferometers, an oscillation in the phase difference as function of the azimuth of the set-up is observed. With regard to the measurement accuracy, this oscillation is constant. 
Analysis of the signals of the interferometers show that the fiber optical asymmetric Mach-Zehnder interferometer built, is unable to detect a sidereal variation in the anisotropy of the velocity of light at the Earth surface. The origin of the observed oscillations in the signal upon rotation in the laboratory are not identified with certainty and are subject to further future exploration. 
\\
It is questionable that calculations of travel-time differences could yield correct first order predictions for phase difference in interferometers.


\begin{thebibliography}{1}
\bibitem{Smooth1977}Smoot G F and Gorenstein M V  and Muller R A,  Physical Review Letters {\textbf 39} 14 (1977) p898
\bibitem{DiSalle2008} DiSalle  R, "Space and Time: Inertial Frames", The Stanford Encyclopedia of Philosophy (Fall 2008 Edition), Edward N. Zalta (ed.), http://plato.stanford.edu/archives/ fall2008/entries/spacetime-iframes/
\bibitem{Cahill2003}Cahill R T and Kitto K, Apeiron {\textbf 10} 2 (2003) p104-117
\bibitem{Consoli2004}Consoli M and Costanzo E, Physics Letters A {\textbf 333} (2004) p355–363
\bibitem{Miller1933}Miller D C, Reviews of Modern Physics {\textbf 5} (1933) p203-242
\bibitem{Piccard1926} Piccard A and Stahel E, Comptes Rendus {\textbf 183} (1926) p420-421
\bibitem{Piccard1928} Piccard A and Stahel E, Comptes Rendus {\textbf 184} (1928) p152
\bibitem{Illingworth1927}Illingworth K K, Physical Review {\textbf 30} (1927) p692-696
\bibitem{Joos1930}Joos G, Annaler der Physik S 5 {\textbf 7} 4 (1930) p385-407
\bibitem{Trimmer1973}Trimmer W S N and Baierlein R F and Faller J E and Hill H A, Physical Review D {\textbf 8} 10 (1973) p3321-3326
\bibitem{DeHaan2009} de Haan V O, Canadian Journal of Physics, accepted 
\bibitem{Edlen1967} Edl\'{e}n B, Metrologia {\textbf 2} 2 (1966) p12-80
\bibitem{Munera1998}M\'{u}nera H A, Apeiron {\textbf 5} No 1-2 (1998) p37-53
\bibitem{Conference1927}Conference on the Michelson-Morley Experiment held at the Mount Wilson Observatory Pasadena, California, February 4 and 5, 1927, Astrophysical Journal {\textbf 68} 5 (1928) p341-402
\bibitem{Christov2006}Christov C I, Progress in Physics {\textbf 3} (2006) p55-59
\end{thebibliography}
\end{document}